\documentclass[prb,aps,groupedaddress,amsmath,twocolumn,floatfix,showpacs]{revtex4}

\usepackage[dvips]{graphicx}
\usepackage{dcolumn}
\usepackage{bm}
\usepackage{epsfig}
\usepackage{amsmath}

\newcommand{\ud}{\,\mathrm{d}}

\begin{document}

\title{Exact quantum Monte Carlo study of
one dimensional trapped fermions with attractive contact interactions}

\author{Michele Casula}
\affiliation{ Department of Physics, University of Illinois at Urbana-Champaign,
1110 W. Green St, Urbana, IL 61801, USA}

\author{D. M. Ceperley}
\affiliation{ Department of Physics, University of Illinois at Urbana-Champaign,
1110 W. Green St, Urbana, IL 61801, USA}

\author{Erich J. Mueller} 
\affiliation{ Laboratory of Atomic and Solid State Physics, Cornell University, 
Ithaca, New York 14853, USA}

\date{\today}

\begin{abstract}
Using exact continuous quantum Monte Carlo techniques, we
study the zero and finite temperature properties of a system of harmonically trapped 
one dimensional spin 1/2 fermions with short range interactions.
Motivated by experimental searches for modulated Fulde-Ferrel-Larkin-Ovchinikov states, we 
systematically examine the impact of a spin imbalance on the density profiles.
We quantify the accuracy of the Thomas-Fermi approximation, finding that 
for sufficiently large particle numbers ($N \gtrsim 100$) 
it quantitatively reproduces most features of the exact density profile.  
The Thomas-Fermi approximation fails to capture small Friedel-like spin 
and density oscillations and overestimates the size of the fully paired
region in the outer shell of the trap.
Based on our results, we suggest a range of experimentally
tunable parameters to maximize the visibility of the double
shell structure of the system and the
Fulde-Ferrel-Larkin-Ovchinikov state in the one dimensional
harmonic trap.
Furthermore, we analyze the 
fingerprints of the attractive contact interactions in the
features of the momentum and pair momentum distributions.
\end{abstract}


\maketitle

\section{Introduction}

Mapping the phase diagram of fermions with attractive interactions
and spin imbalance is a challenging problem, both from the experimental
and the theoretical side. For an unpolarized system,
an effective attractive interaction leads to the formation of Cooper pairs
carrying zero momentum.\cite{cooper} 
However in the presence of spin imbalance, where the different spin components have
their Fermi energies shifted from one another, the pairing mechanism changes.
Depending on parameters, different theories apply.
The Sarma or breach-pair theory, which is stable in the presence of long range interactions, suggests
a coherent superposition of Bardeen-Cooper-Schrieffer
(BCS) pairs and normal fermions,\cite{sarma,liu,forbes,gubbels}
while in other regimes one finds Fermi surface deformations\cite{muther,sedrakian}
or a coexistence of normal fluid and superfluid.\cite{bedaque}
One of the most interesting possibilities is the
Fulde-Ferrel-Larkin-Ovchinikov (FFLO) states,\cite{fulde,larkin,casalbuoni}
which reconcile spin imbalance and superconductivity
by pairing particles whose momentum does not sum to zero.  For example,
in the FF state, $\uparrow$-spin fermions with momentum $k$ are paired 
with $\downarrow$-spin fermions with momentum $-k+q$, resulting in a Cooper pair with momentum $q$.  
In the LO theory, the Cooper pairs carry no net current, but are in superpositions of momentum states, 
leading to an order parameter that oscillates in space; the excess spin polarization resides 
in the nodes of the order parameter. Analogs of the latter state are found in one dimension (1D). 
We will use an exact quantum Monte-Carlo procedure to study a gas of 
spin imbalanced harmonically trapped 1D fermions, 
with a goal of identifying the signatures of FFLO physics which will be seen in cold atom experiments.

Although not completely transparent, the best evidence for the FFLO state 
in solid-state systems comes from  layered organic superconductors\cite{singleton}
and heavy-fermion materials.\cite{radovan,miclea}
However, recent progress in optical lattice experiments with cold atoms has opened the avenue
to study spin imbalanced fermions with attractive interactions in a clean and direct way.
\cite{zwierlein,zwierlein2,shin,schunck,partridge,partridge2}
Unfortunately, for a bulk three dimensional gas of atoms with short range interactions, 
the  FFLO state is found for a very narrow range of chemical potentials,\cite{parish} 
implying that only a small volume of a trapped atomic cloud would be in this state.  
Hence attention has been shifting to one dimension, where an analog 
of the FFLO state has a very large region of stability.\cite{liu_hu_drumm_fflo} 

Since there is no true off-diagonal long-range order in a strict 1D system, 
the 1D FFLO state is characterized by an algebraic quasi-1D order.
Although the thermodynamics of a uniform 1D Fermi gas with contact interactions can be 
found exactly via the Bethe ansatz at zero\cite{gaudin,takahashi} and finite
temperature,\cite{takahashi_temp,guan} the characterization of the ground state
with spin imbalance is not easy: many correlation functions
are unknown due to the complicated structure of the Bethe ansatz.  
Other properties are well known: for example, Yang firmly established 
the existence of the FFLO state through a bosonization technique, showing that it
emerges via a continuous transition from the BCS state.\cite{jang} 
Further, numerical methods have shown that the FFLO
is the ground state of the Hamiltonian on a lattice
as soon as the system is polarized \cite{batrouni,rizzi,luscher}.

Experimentally, an array of quasi one dimensional tubes can be created at the
intersection of two orthogonal standing waves generated by laser beams.\cite{paredes,moritz,moritz2}
Along each tube there would be no lattice, but the particles will be confined by a longitudinal
harmonic potential, coming from the spatial profile of the lattice lasers, 
and possibly from some additional potential.
Recently Orso\cite{orso} and Liu \emph{at al.}\cite{liu_hu_drumm_prl}
studied the properties of an atomic cloud in such a 1D harmonic trap.
By using a Thomas Fermi (TF) approximation based on the Bethe ansatz solution
of the homogeneous Hamiltonian, they found that in the trap the system phase separates. 
The inner shell of the tube is always a partially polarized FFLO state, while the outer shell can be
either a fully paired BCS or a fully polarized normal state.  
Our calculations are designed to go beyond the approximations inherent in those calculations, 
and accurately quantify the properties of these 1D clouds.

Other researchers have also attempted to investigate the accuracy of the TF approximation.  
For example, Xianlong and Asgari developed an exchange-correlation functional
to perform density functional theory (DFT) calculations
with the local density approximation (LDA).\cite{xianlong}
While revealing, those calculations are only approximate.
In a different direction, several researchers have used 
the density matrix renormalization group (DRMG) to study a related system, 
where there is a 1D lattice along the tube.\cite{feiguin,tezuka,capelle}
One needs to be cautious about applying those results to the continuous system as the
underlying lattice in the model could introduce effects of incommensurability in the density profiles.\cite{buchler,molina}
The continuous limit is difficult to access due to the large number of sites 
required in the DMRG evaluation for an extremely dilute system. 

We will also consider how temperature affects the density profiles. 
Since the superfluid critical temperature is zero in 1D, thermal effects could be quite dramatic, 
and may not be captured by mean field theories.\cite{liu_hu_drumm_finitet} 
Experiments will inevitably be performed at finite temperature, and it is imperative 
to know to what extent finite temperature will obscure the clear delineation of phases 
which are expected in a trapped zero temperature gas.

In this paper, we present exact results for 1D fermions subject
to a zero-range attractive interaction and a longitudinal
harmonic confinement.  We resolve the dependence on the
spin imbalance, effective coupling, and temperature, within a
continuous quantum Monte Carlo (QMC) framework. We use 
diffusion Monte Carlo (DMC)\cite{reynolds:5593} for
zero temperature ground state calculations and path integral
Monte Carlo (PIMC)\cite{ceperley_pimc} for finite temperature,
which are ideal numerical tools to find the properties of 1D Hamiltonians, as they do
not suffer from the ``sign problem'' which affects QMC
simulations in higher dimensions.\cite{ceperley91} In 1D they
give unbiased energies, with their accuracy only limited by statistical noise. 
We correct other potential errors in the DMC algorithm,
such as the time step $\tau$, population bias, and mixed
estimate errors, by appropriate
extrapolation and forward walking
techniques.\cite{calandra:1998} The only systematic bias in the
PIMC algorithm is the time step, which we control by working
with sufficiently small $\tau$.

The paper is organized as follows.
In Sec.~\ref{model} we outline the model Hamiltonian used
in our calculations, and the experimental parameters of the system.
Sec.~\ref{methods} includes a description of the QMC methods
with an emphasis on the features required to simulate 1D fermions
interacting with a zero-range potential. In Sec.~\ref{results}
we present our results and compare them with the TF findings.
In particular, we focus our attention
to the finite size and temperature effects on the phase boundaries.
We conclude in Sec.~\ref{conclusions} with some highlights and remarks.

\section{Model}
\label{model}

The Hamiltonian which describes 1D fermions in a trap and interacting
with an attractive zero-range potential is
\begin{equation}
H = -\frac{\hbar^2}{2m} \sum_{i=1}^N \frac{\partial^2}{\partial x_i^2}
- g \sum_{i=1}^{N^\uparrow} \sum_{j=1}^{N^\downarrow} \delta(x_i-x_j)
+ \frac{1}{2} m \omega_z^2 \sum_{i=1}^N x_i^2,
\label{hamiltonian}
\end{equation}
where $N=N^\uparrow+N^\downarrow$ is the total number of particles with mass $m$,
$g(>0)$ is the strength of the contact interaction, and
$\omega_z$ is the harmonic frequency of the longitudinal trap.
The effective 1D Hamiltonian
in Eq.~\ref{hamiltonian} is a very good representation
of a single tube in the optical array
provided that $N \hbar \omega_z \ll \hbar \omega_\perp$
and $k_B T \ll \hbar \omega_\perp$, where $\omega_\perp$ is the
frequency of the transverse harmonic confinement. The above
requirements are met in the experimental setup by
properly tuning the height of the lattice depth,
which sets the value of $\omega_\perp$.
In such a 1D geometry with just one open channel,
the interactions can be modeled by as
$V(x)=- g \delta(x)$
 with $g$ given by\cite{olshanii,bergeman}
\begin{equation}
g=\frac{2 \hbar^2 a_{3D}}{m a^2_\perp} \frac{1}{A a_{3D} / a_\perp -1},
\label{scattering}
\end{equation}
where $a_{3D}$ is the 3D $s$-wave scattering length,
$a_\perp = \sqrt{\hbar/m \omega_\perp}$ is the transverse oscillator length, and 
$A= -\zeta(1/2)/\sqrt{2} \simeq 1.0326$, where  $\zeta$ is the Riemann zeta function.
Here and throughout the paper  we have used a sign convention where $g>0$ is attractive.
An ongoing experiment at Rice University\cite{randy} is using $N \sim 100-300$
of $^6\textrm{Li}$ atoms per tube, the actual value depending on the position
of the tube in the array.
Typical values for the optical lattice are
$\omega_\perp = 2 \pi~ 156~ kHz$, $\omega_z = 2 \pi ~200 ~Hz$,
while $g$ will be tuned by changing $a_{3D}$.
The typical temperatures are around $T \sim 0.05 ~T^\uparrow_F$,
with $T^\sigma_F = N^\sigma \hbar \omega_z / k_B$
the Fermi temperature of fermions with spin $\sigma$.
Throughout the paper we will use $m=\hbar=\omega_z=k_B=1$,
i.e. the length will be measured in units of the harmonic
oscillator length $\sqrt{\hbar/m \omega_z}$, and
the energy and temperature in units
of the level spacing $\hbar \omega_z$.
In these units the energy for $g=0$ is $E=(N_\uparrow^2+N_\downarrow^2)/2$ and
the binding energy of a pair is $g^2/4$.

\section{Methods}
\label{methods}

The ground state QMC techniques used in this work are described
in Subsec.~\ref{zero_temperature_QMC}, while the finite
temperature PIMC method is reported in
Subsec.~\ref{finite_temperature}. In order to compare our
results with the local density approximation, we computed also
the TF density profile ($n(x)=n^\uparrow(x)+n^\downarrow(x)$)
and spin density profile ($s(x)=n^\uparrow(x)-n^\downarrow(x)$)
of the system, which are solutions of:
\begin{equation}
\label{thomas_fermi}
\mu_{\textrm{hom},\sigma}[n(x),s(x)] + \frac{1}{2} x^2 = \mu_\sigma \qquad  \sigma=\uparrow,\downarrow,
\end{equation}
where $\mu_\sigma$ is the chemical potential of the species with spin $\sigma$, and
$\mu_{\textrm{hom},\sigma}$ is the local chemical potential found from the Gaudin
exact solution\cite{gaudin,takahashi} of the homogeneous model. The Thomas-Fermi
results have been widely discussed elsewhere,
and we refer the reader to previous works\cite{orso,liu_hu_drumm_prl,liu_hu_drumm_fflo,xianlong}
for complete details. Here, we mention that
Eq.~\ref{thomas_fermi} can be rescaled into dimensionless variables
with coupling strength controlled by $4 N / g^2$. Therefore, the effective
coupling constant of the trapped system depends also on the number of particles
and scales as $g/\sqrt{N}$.

\subsection{Ground state QMC}
\label{zero_temperature_QMC} To study the ground state
properties of the system, we employed variational and
diffusion quantum Monte Carlo methods. Their basic ingredient
is the trial wave function $\Psi_T$, a good approximation to
the ground state. We used the form:
\begin{equation}
\Psi_T = D^\uparrow D^\downarrow \exp(-U),
\end{equation}
where $U$ is a bosonic function and $D^\sigma$ is the
antisymmetrized product of Hermite polynomials of $i$-th order
$H_i$ for particles with spin $\sigma$. 
$D^\sigma$ can be written as determinant of a Van der Monde matrix:
\begin{equation}
D^\sigma(x_1, \ldots, x_{N^\sigma}) =
\det(H_i(x_j)) =  \prod_{\substack{1 \le i < j \le N^\sigma}}(x_i-x_j).
\label{hermite_van_der_monde}
\end{equation}
The product in the Eq.~\ref{hermite_van_der_monde} is convenient, as it can be
evaluated in order $N_\sigma^2$ operations. The Hermite basis
set yields an exact variational wavefunction in the limit of vanishing inter-particle
interactions. The Slater term $D^\uparrow D^\downarrow$
determines the nodes of the variational wave function. In 1D,
the exact nodes of the ground state are defined by the
coalescence conditions between two like-spin particles (i.e.
$x_i=x_j$ with $\sigma_i=\sigma_j$).  This result holds even for an interacting system, and 
consequently
there is no sign problem in the diffusion Monte Carlo
calculations.\cite{ceperley91}

The bosonic function $U$ is real and symmetric under
permutation of two fermions, and is the sum of one-body ($U_{1b}$), 
two-body ($U_{2b}$), and three-body ($U_{3b}$) correlations. 
The one-body part reads:
\begin{equation}
U_{1b}= \sum_{i=1}^N f^{\sigma_i} (x_i),
\end{equation}
where $\sigma_i$ is the spin of the $i$-th particle, and 
$f^\sigma(x)$ is written in a Pad\'e form which goes as $x^2/2$
at large $x$ so to assure the right asymptotic behavior given
by the harmonic confinement. In the absence of inter-particle
interactions, $f^\sigma(x)=x^2/2$  is the exact solution.

The two-body (Jastrow) factor is defined as:
\begin{equation}
U_{2b}=\sum_{\substack{1 \le i<j \le N}} u^{\sigma_i \sigma_j}\left( x_i - x_j \right),
\end{equation}
with $u^{\sigma_i \sigma_j}$ pair functions chosen to fulfill the cusp
conditions set by the contact interactions of the Hamiltonian
in Eq.~\ref{hamiltonian}. In the case of two unlike-spin
particles interacting with a delta function potential, the
exact solution is given by
$\Psi(x^\uparrow,x^\downarrow)=\exp(-\frac{g}{2} | x^\uparrow -
x^\downarrow |)$; note the cusp at $x^\uparrow = x^\downarrow$.
The exact many-body function will have the same cusp between atoms with different spin.
We use the following pair functions:
\begin{eqnarray}
u^{\uparrow \downarrow}(x) & = & -\frac{g}{2 \lambda}
\exp\left(-\lambda |x|\right),
\label{u_cusp}\\
u^{\sigma \sigma}(x) & = & \alpha_{\sigma \sigma}
\frac{\exp(-\beta_{\sigma \sigma} |x|)}{1+ \exp(- 2
\beta_{\sigma \sigma} |x|)},\qquad  \sigma \in \{\uparrow,\downarrow\},
\label{u_cuspless}
\end{eqnarray}
where $\lambda$, $\alpha_{\sigma \sigma}$,
and $\beta_{\sigma \sigma}$ are variational parameters, and $g$ is the 
strength of the contact interaction.

The confining potential leads to inhomogeneities that make it
convenient to include an inhomogeneous  Jastrow function:
\begin{equation}
U^{\textrm{non-homo}}_{2b}=\sum_{1 \le i < j \le N}
u^{\sigma_i \sigma_j}(x_i - x_j)~
f^{\sigma_i \sigma_j}(x_i + x_j),
\end{equation}
where the pair function $u^{\sigma_i \sigma_j}$ is modulated
by the function $f^{\sigma_i \sigma_j}$ which depends on the center of mass
of the particles $x_i$ and $x_j$.
The analytic form of $u^{\sigma \sigma^\prime}$ 
is the same as in Eq.~\ref{u_cuspless} also for $\sigma \ne \sigma^\prime$, 
while $f^{\sigma \sigma^\prime}$ is chosen to be:
\begin{equation}
f^{\sigma \sigma^\prime}(x)=1 + \gamma_{\sigma \sigma^\prime}
\frac{\exp(-\delta_{\sigma \sigma^\prime} x)}{1+ \exp(- 2 \delta_{\sigma \sigma^\prime} x)},
\end{equation}
where $\gamma_{\sigma \sigma^\prime}$ controls the inhomogeneity.
For $\gamma_{\sigma \sigma^\prime}=0$ one recovers the homogeneous case.

Finally, we employed a three-body term:
\begin{equation}
U_{3b}=\sum_{\substack{i,j,k=1}}^N 
u^{\sigma_i \sigma_j}(x_i - x_j)~
u^{\sigma_j \sigma_k}(x_j - x_k),
\end{equation}
where also in this case the functional form of $u$ is the same
as in Eq.~\ref{u_cuspless}. We found this term important
in the intermediate and strong coupling regime,
when the local energy is dominated by the inter-particle potential.

Our variational ansatz includes about 30 parameters $\{p_i\}$,
optimized by means of the stochastic reconfiguration
algorithm,\cite{sorella:024512,casula:7110} which minimizes the
total energy $E_T$ of the wave function with an iterative
procedure based on the value of the forces, $-\partial E_T /
\partial p_i$, acting on the parameters at each step. The
optimal variational wave function is then projected to the
ground state by using the diffusion Monte Carlo method
(DMC)\cite{reynolds:5593}, which provides an unbiased estimate
of the ground state energy.
The physical observables, such as the charge and spin density
profiles, are computed with the forward walking propagation of
the mixed DMC distribution.\cite{liu74,calandra:1998}

\subsection{Finite temperature QMC}
\label{finite_temperature}

The finite temperature calculations are performed by using the
path integral Monte Carlo (PIMC) method\cite{ceperley_pimc}
which is based on sampling the thermal density matrix
$\rho(R,R^\prime;\beta)$, with $\beta=1/k_BT$, $T$ the temperature, 
by means of a mapping of the quantum system into a classical system of
polymers or Feynman\cite{feynman}``paths''.
$R \equiv \{x_1,\ldots,x_N\}$ and $R^\prime \equiv \{y_1,\ldots,y_N\}$ 
are all-particle configurations.
During the Monte Carlo random walk the polymers are distributed
according to the probability
\begin{equation}
\exp \left( -\sum_{m=1}^M S_m \right),
\label{prob_action}
\end{equation}
where the integration from $0$ to $\beta$ has been discretized
into $M$ ``time slices'' of length $\tau=\beta/M$,
$S_m=-\log\left[\rho(R_{m-1},R_m;\tau) \right]$ is the action,
and $R_0=R_M$. An accurate approximation of
$\rho(R,R^\prime;\tau)$ for the system under consideration is
important to make the PIMC calculations feasible and efficient.
In the case of the Hamiltonian in Eq.~\ref{hamiltonian},
$\rho(R,R^\prime;\tau)$ can be written using a Trotter expansion
as a product of four factors: the non-interacting ($\rho_0$),
the harmonic ($\rho_\textrm{harm}$), the interacting
($\rho_\textrm{int}$), and the fermionic
($\rho_\textrm{Fermi}$). The non-interacting part corresponds to the propagator for free particles
\begin{equation}
\rho_0(R,R;\tau)=(2\pi \tau)^{-N/2} \exp\left\{-(R-R')^2/2
\tau\right\}. \label{non-interacting}
\end{equation}
We use the primitive approximation\cite{ceperley_pimc} to
include the effect of the harmonic external potential:
\begin{equation}
\rho_\textrm{harm}(R,R';\tau)=\exp\left\{-\frac{\tau}{4}\left(R^2
+ R'^2 \right)\right\}. \label{harmonic}
\end{equation}
To account for the interactions we make use of the exact action for
two particles in free space interacting with an attractive
contact potential \cite{gaveau}. In terms of the relative
distances of the pair ($x_\textrm{rel} \equiv x_1-x_2$,
$y_\textrm{rel} \equiv y_1-y_2$), the resulting propagator is:
\begin{eqnarray}
\rho_\delta(x_\textrm{rel},y_\textrm{rel};\tau)= & & 1 +
\sqrt{\pi} s \exp\left\{ s(s-2r) \right . \nonumber \\ 
 & & \left . + z^2(x_\textrm{rel}-y_\textrm{rel})^2\right\}
\textrm{erfc}\left(r-s\right),
\label{2body_delta}
\end{eqnarray}
where $s=\frac{g}{2} \sqrt{\tau}$, $z=1/\sqrt{4 \tau}$, $r=z
\left( |x_\textrm{rel}|+|y_\textrm{rel}| \right) $.
From this propagator, one  obtains the
many-body action by means of the pair-product
approximation\cite{ceperley_pimc}
\begin{equation}
\rho_\textrm{int}(R,R^\prime;\tau) = \prod_{i=1}^{N^\uparrow} \prod_{j=1}^{N^\downarrow} \rho_\delta(x_i-x_j,y_i-y_j;\tau),
\label{delta}
\end{equation}
where the product runs over all unlike-spin pairs.  This, as the other approximations, becomes exact as $\tau\to0$.

To include Fermi statistics into the PIMC method
\cite{ceperley_fn} the paths are constrained by the nodal
regions, which act like barriers. Since the nodes are exactly
known in 1D, and they do not depend on $\beta$, the restricted
path integral formalism gives the exact result. An approximate
fermion action is given by the image approximation:
\begin{equation}
\rho^\sigma_\textrm{Fermi}(R,R^\prime;\tau)=\prod_{i=2}^{N^\sigma}
\left(1 - \exp\left\{ \frac{(x_{i-1}-x_i)
(y_{i-1}-y_i)}{2 \tau} \right\}\right),
\label{fermi}
\end{equation}
for each spin sector. It includes the nodes between neighboring
particles as zero's of the density matrix, and becomes exact in
the $\tau=0$ limit.  We assume that the like spins are ordered
$x_{i-1} < x_i$ for $1<i\leq N_{\sigma}$.

The probability distribution in Eq.~\ref{prob_action}, based on
the the density matrix $\rho_\textrm{int}(R,R^\prime;\tau)$ defined
above, is sampled via a generalized Metropolis algorithm. The
paths are displaced with two types of moves: a multi-level
bisection (level two and three have been used here) and a
global move, which attempts to displace an entire polymer at
once. When the convergence is reached, the thermal expectation
values are computed as described in
Ref.~\onlinecite{ceperley_pimc}.

\section{Results}
\label{results}

\begin{figure}[h]
\centering
\includegraphics[angle=0, width=\columnwidth]{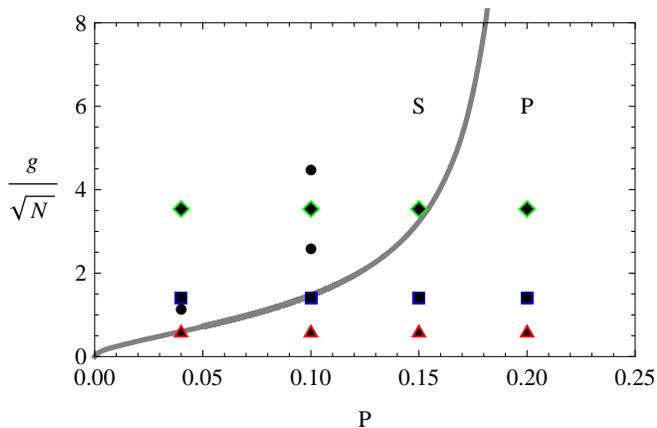}
\caption{(Color online) 
Thomas-Fermi phase diagram in the polarization-coupling plane 
for 1D harmonically trapped cloud of two-component
fermions, interacting through a contact potential $V(x_i-x_j)=-g \delta(x_i-x_j)$.
The number of particles is $N$, the polarization $P$, and we use units where
the harmonic confinement energy and length scales are set equal to unity. 
The gray thick line separates the phase region where the outer shell of the cloud
is fully paired (S) from the region with fully polarized outer shell (P).
The points locate our QMC simulations on the TF phase diagram. 
Triangles (red outline), squares (blue outline), and diamonds (green outline) 
refer to DMC data reported in Fig.~\ref{all_n200},
panels (a), (b), and (c) respectively. 
The two circles at $P=0.10$ depict 
systems with the same $g=20$ but different sizes ($N=60$ and $20$),
displayed in Fig.~\ref{polvsN}. 
The circle at $P=0.04$ represents
the system with $N=200$ and $g=16$, 
whose ground-state DMC polarization is reported in Fig.~\ref{polvsg}.
At the same polarization, the triangle, circle and square points
have been studied also by PIMC, which yielded 
temperature dependent polarizations reported in 
Figs.~\ref{pol_pimc_g8}, \ref{pol_pimc_g16}, and \ref{pol_pimc_g20} respectively.
}
\label{tfphase}
\end{figure}

The QMC results shown in this section characterize the ground
state and finite temperature properties of the system. We compute the density
$n(x)$, spin density $s(x)$, and local polarization
($p(x)=s(x)/n(x)$) profiles. The latter quantity is extremely
useful, as it locates the paired ($p=0$), partially polarized ($0<p<1$), and fully
polarized ($p=1$) shells. For sufficiently high densities, 
the center of the cloud is always partially polarized, but the edge may be paired or fully polarized.  
The Thomas-Fermi approximation yields the phase diagram shown in Fig.~\ref{tfphase}, 
which shows the expected boundary between these two regimes.
We decided to predominantly study the system with
$N=200$, since it is roughly the average number of fermions per
tube in typical experiments. As marked on the phase diagram,
we present results for a range of interaction strengths and polarizations.  
In our numerical profiles we are particularly interested in how the visibility of the interface 
between different regions is influenced by finite particle number and finite temperature.  
By analyzing the experimental data within a Thomas-Fermi approximations 
one could use the locations of these boundaries to map out the phase diagram.  
We find that for $N \gtrsim 100$ and $T\lesssim 0.05 T_F$ this Thomas-Fermi analysis is reasonable, 
but for fewer particles or higher temperatures one would need to use a much more sophisticated analysis 
of the experimental data to learn about the bulk zero temperature phase diagram.
We report our results for the ground state and
the finite temperature calculations in
Secs.~\ref{zero_temperature_results} and
\ref{finite_temperature_results} respectively.

\subsection{Ground state}
\label{zero_temperature_results}

\begin{figure}[h!] 
\centering
\includegraphics[angle=0, height=15cm]{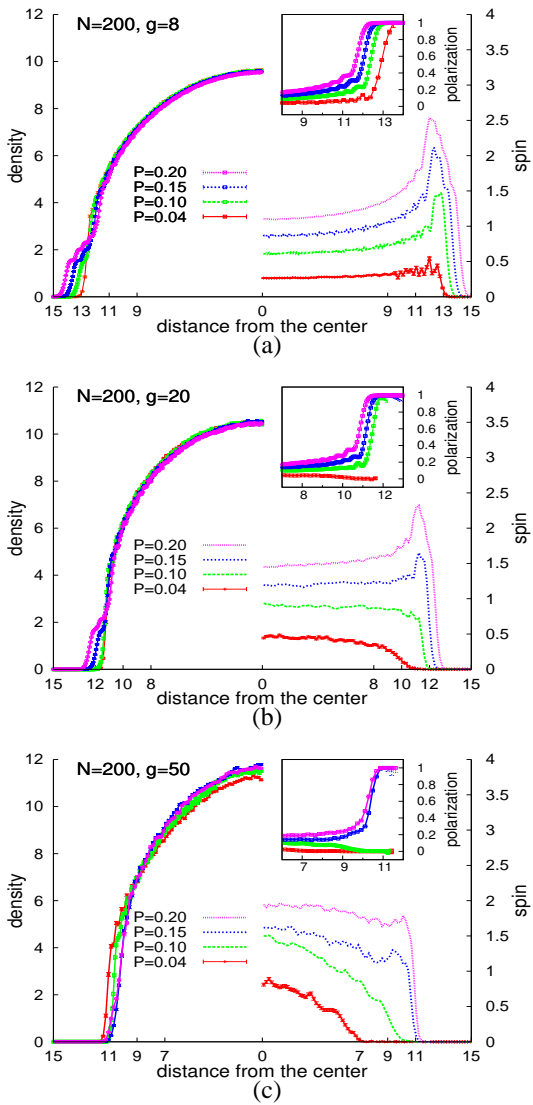}
\caption{(Color online) 
Density $n$, spin density $s$ and local polarization $p=s/n$ profiles
for a cloud with $N=200$.
The panel (a) shows the QMC results for
$g=8$ and four different polarizations $P$, while panels (b) and (c) display
the QMC density profiles for $g=20$ and $50$ respectively,
with the same set of polarizations.
The spin density is reported at the right side of the plot, while
the density is on the left. The inset shows the polarization 
at the edge of the cloud, where
the crossover from the FFLO core to the outer shell occurs.
At $g=8$ the edge of the cloud is a completely polarized
normal gas for all polarizations, whereas
for $g=20$ and the lowest polarization $P$ 
the edge is in the fully paired state.
At even stronger coupling
($g=50$), both $P=4\%$ and $P=10\%$ give fully paired outer shells.
Finally, note how the strength of the interaction shrinks the cloud, while
up to $P=20\%$ the polarization has a little impact on its spread. 
At $g=50$ the equilibration of the system begins to break down in the MC sampling, 
as particles are unable
to easily cross each other, being the pairs strongly bound.  
This breakdown accentuates the MC noise in the results.}
\label{all_n200}
\end{figure}

In Fig.~\ref{all_n200} we report on the density, spin density, and local polarizations 
of $N=200$ particles with various polarizations $P$ and interaction strengths $g$.  
From the TF results in Fig.~\ref{tfphase}, 
one expects that the edge of the cloud will be fully polarized 
for $(g,P)=(8,0.2), (8,0.15), (8,0.1), (20, 0.2), (20,0.15)$.  
Similarly, the edge should be fully paired for $(g,P)=(20,0.04),(50,0.1),(50,0.04)$.  
The others: $(g,P)=(8,0.04),(20,0.1),(50,0.15)$ are near-critical, 
and one expects that the FFLO state should extend nearly to the edge of the cloud.  
The results in Fig.~\ref{all_n200} are consistent with these expectations.  
For the TF near-critical points, we found that all of them feature fully polarized wings.
When the outer shell of the cloud is polarized, 
the spin density has a peak at the edge, 
which is replaced by a rapid fall-off when the edge is fully paired.
Detecting the fully paired shell will be easiest at strong coupling, 
where the size of the outer paired region can be large.

Notice that the $g=50$ spin
polarization plotted in Fig.~\ref{all_n200}(c) is much more
noisy than for weaker interactions. 
Indeed, from the computational point of view a problem
of ergodicity starts affecting the Monte Carlo (MC) sampling.
As the coupling increases, one needs to run longer to
equilibrate the sample and have good statistics in the
expectation values, because sometimes the evolution gets stuck
in a ``persistent'' configuration. In this regime, it is
difficult for unpaired fermions to cross a strongly bound pair
due to the Pauli principle. We note that it is possible that
the same physics will affect also the dynamics of the $^6\textrm{Li}$
atoms in the experiment. The physical origin of the ergodicity
problem in the MC evolution, namely the formation of strongly
bound pairs, could also cause  a slow equilibration in the
experimental setup. A coupling between $g=20$ and $g=50$ is a good balance between
the visibility of the fully paired region and the ergodicity of
the pairs in the tubes.

\begin{figure}[h] 
\centering
\includegraphics[angle=0, width=\columnwidth]{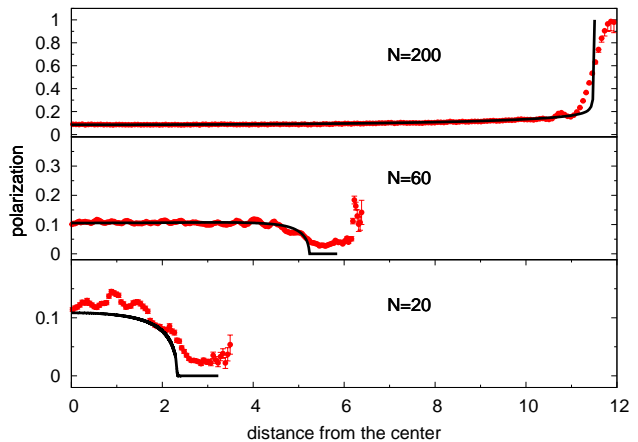}
\caption{Local polarization profiles for $g=20$, $P=10\%$,
and different number of particles $N$. The TF polarization is drawn (thick black line)
for comparison. The TF approximation gives fully paired wings with zero polarization
for $N=20$ and $60$, while the QMC polarization is always positive. The agreement between 
QMC and TF is better for the largest system ($N=200$), where the outer shell is fully
polarized in both methods.
}
\label{polvsN}
\end{figure}

From the analysis presented above, it seems that the TF mean
field phase diagram is a good approximation for the couplings
and polarizations studied so far. It is clear that the TF
approximation is not able to reproduce the short length-scale charge and spin
fluctuations, since the local density approximation assumes a
slow varying density profiles, as already pointed out in
previous studies.\cite{mueller,xianlong} However, the TF
boundaries of the phase separated states are in a quite good
agreement with our exact QMC calculations. In order to better
understand this agreement, we study the dependence of the local
polarization on the number of particles $N$ and coupling $g$,
and compare our QMC results with the TF predictions. The local
polarization for different system sizes ($N=20$, $60$, $200$)
with $g=20$ and $P=10\%$ is plotted in Fig.~\ref{polvsN}. At
$N=20$ and $60$, there are qualitative disagreements between
the QMC and TF methods, which disappears at $N=200$. Indeed for
small systems our QMC calculations give partially polarized
wings, while they are fully paired in the TF approximation.
This can be due to a proximity effect not accounted for in the
completely local TF approximation. 
We speculate that the FFLO state partially extends into the
wings, making them partially polarized.  We cannot
confirm this picture since we did not study how the FFLO order
parameter depends on position. For $N=200$ the wings become
fully polarized, as the effective coupling is reduced, and the
polarization profile turns out to be in agreement with the TF
results, although the transition from the FFLO to the fully
polarized state is much smoother in our QMC calculations.
The boundaries between phase separated states are smeared 
out over a length-scale of order the atomic spacing. For small
systems the very concept of phase separation begins to break down.

\begin{figure}[h] 
\centering
\includegraphics[angle=0, width=\columnwidth]{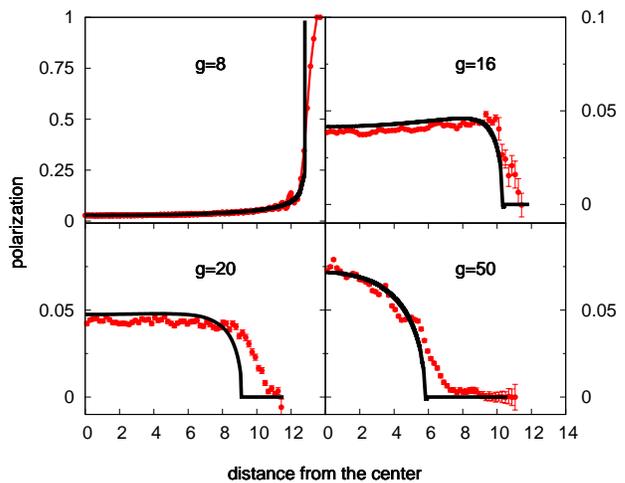}
\caption{Local polarization profiles for $N=200$, $P=4\%$,
at various couplings $g$. Also the TF polarization is drawn (thick black line)
for comparison. The QMC results show a crossover to fully paired wings for $g \ge 16$,
which is much smoother than the corresponding TF transition. Moreover,
the QMC simulations give a larger partially polarized shell in the center of the trap.
}
\label{polvsg}
\end{figure}

We check the agreement between the QMC and TF methods for large
number of particles ($N=200$) in a more systematic way by
studying the dependence of the polarization on $g$ with
$P=4\%$. These results are plotted in Fig.~\ref{polvsg} for
$g=8$, $16$, $20$, and $50$. The TF polarization profile
follows quite closely the DMC points. Both theories predict a
``transition'' from a fully polarized shell to a fully paired
one between $g=8$ and $16$. Apart from fluctuations and
statistical noise on the points, the only significant
difference in the comparison is the size of the fully paired
region which is smaller in the QMC results, where the FFLO
state seems to extend farther from the center of the trap. This
is a common feature of our calculations. The TF approximation
always overestimates the size of the fully paired shell.
Apparently, this is a drawback of the local density
approximation, which makes the effective attractive interaction
stronger.\cite{xianlong_lda}

We complete the comparison between the TF and QMC results by also computing
the energetics of $g=20$ with
different polarizations ($P=0\%$, $10\%$, $20\%$)
and system sizes ($N=20$, $200$), reported in Tab.~\ref{total_energies}. 
The agreement in the ground state energies per particle between
the two methods is remarkable, even 
at strong coupling (small $N$).
This means that the total energy is insensitive to the local 
differences seen in the shell structure of the cloud, 
interpreted as finite size or edge effects.
It also highlights the importance of an accurate treatment of  
local correlation effects, which can be missed in the TF approximation, 
and lead to subtle changes in the phase boundaries, whereas the 
overall trend is correctly captured by the mean field method, even at
a quantitative level.

\begin{table}[!htp]
\caption{\label{total_energies}
Ground state energies per particle
computed by means of DMC and TF, for $g=20$
at different polarizations ($P=0\%$, $10\%$, and $20\%$)
and system sizes ($N=20$, and $200$).}
\begin{ruledtabular}
\begin{tabular}{l l d d d d}
\multicolumn{1}{c}{$N$}
& \multicolumn{1}{c}{$P$}
& \multicolumn{1}{c}{TF energy}
& \multicolumn{1}{c}{DMC energy}
 \\
\hline
20 &   0\%   & -47.265   & -47.270(1) \\
20 &  10\%   & -42.550   & -42.538(3) \\
20 &  20\%   & -37.587   & -37.585(2) \\
\hline
200 &  0\%   &  -35.731 & -35.478(10)  \\
200  & 10\%  &  -13.991  & -13.983(3)  \\
200 & 20\%   &  -7.914  & -7.899(3) \\
\end{tabular}
\end{ruledtabular}
\end{table}

\begin{figure}[h] 
\centering
\includegraphics[angle=0, width=\columnwidth]{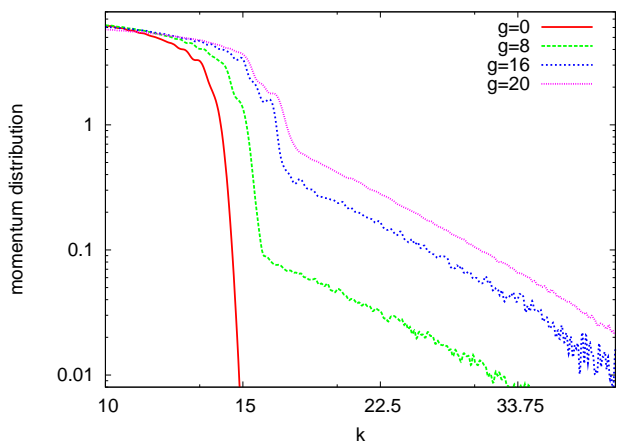}
\caption{(Color online) Momentum distribution for $N=200$, $P=4\%$,
at various couplings $g$, plotted in a log-log scale.
Also the momentum distribution
of the non interacting system is reported. The tails of the interacting $n(k)$
follow a straight lines with a slope compatible with a $1/k^4$ decay.
Momenta are measured in units of $\sqrt{\hbar m \omega_z}$, where $\omega_z$
is the harmonic oscillator frequency along the tube.  For comparison the
characteristic momentum corresponding to the density at the center of the
trap is $k_F=\pi/(2 n)= 16.5$ for $g=20$, while $k_F=14.1$ for the non-interacting system.}
\label{nk_n200_P0.04}
\end{figure}

In previous works,\cite{batrouni,tezuka,feiguin,rizzi} it has
pointed out that the pair momentum distribution function
reveals fingerprints of the FFLO state in the center of the
trap. The momentum and pair momentum distributions can be
extracted in experiments from a statistical analysis of
time-of-flight images of the expanding cloud of atoms when the
confining potential is removed. Here we study the momentum
distribution function defined as
\begin{equation}
n(k) =  \int \ud x \ud y \rho(x,y) e^{i k (x-y)},
\label{momentum_distribution}
\end{equation}
where $\rho(x,y)$ is the one-body density matrix
\begin{equation}
\rho(x,y)= \int \ud x_2 \cdots \ud x_N \Psi(x,x_2,\cdots,x_N) \Psi(y,x_2,\cdots,x_N),
\label{density_matrix}
\end{equation}
and $\Psi$ is the ground state normalized such that
\begin{equation}
\int \ud x \rho(x,x) = N.
\label{norm}
\end{equation}
In the actual DMC calculations, the density matrix in Eq.~\ref{density_matrix}
is computed as follows
\begin{equation}
\rho(x,y) \approx \sum_{i=1}^N \left \langle \delta(x-x_i)
\frac{\Psi_T(\ldots,y,\ldots)}{\Psi_T(\ldots,x_i,\ldots)} \right \rangle_\textrm{MA},
\end{equation}
where $\langle \cdots \rangle_\textrm{MA}$ indicates the MC
averages over $\{x_1,\ldots,x_i,\ldots,x_N\}$ sampled from the
DMC mixed distribution. Although the forward walking technique
for non-local operators is possible, it has rarely been applied
to the evaluation of the momentum distributions.\cite{kalos} 
In our case, we checked that the density matrix
$\rho(x,y)$ computed at the variational MC level is close to
the DMC mixed average (MA), due to the accuracy of the trial wave function
$\Psi_T$. Therefore, our best estimate of
$\rho(x,y)$ is a mixed average with a small bias. In
Fig.~\ref{nk_n200_P0.04} we plot $n(k)$ for different couplings
and $N=200$. We found that the long-range tail of the momentum
distribution decays as a power law as soon as the attractive
contact interaction is switched on, and it is fitted well by
$1/k^4$. This is the same behavior found by Minguzzi \emph{at
al.}\cite{minguzzi:2002} for the tail of the momentum
distribution in a one dimensional gas of hard point-like bosons
(Tonks gas) inside a harmonic trap. The power law decay is due
to the cusp conditions of the wave function, a consequence of
the attractive (or repulsive) contact interactions. The same
decay has been shown recently by Santachiara and Calabrese for
a one dimensional gas of impenetrable anyons.\cite{santachiara}
As one can see in Fig.~\ref{nk_n200_P0.04}, the momentum
distribution does not discriminate between different phase
separated states, but from the height of its tail one can get
information on the effective interaction strength in the trap.

\begin{figure}[h] 
\centering
\includegraphics[angle=0, width=\columnwidth]{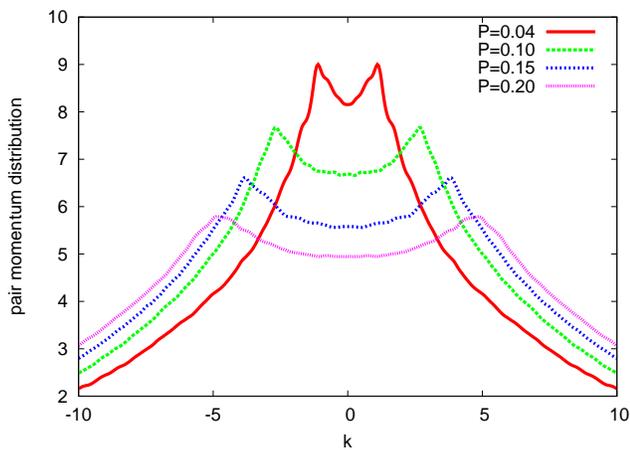}
\caption{(Color online) Pair momentum distribution (Fourier transform of the
pair two-body density matrix in Eq.~\ref{pair_density_matrix}) 
for $g=20$, $N=200$, at various polarizations $P$. 
Momenta are measured in units of $\sqrt{\hbar m \omega_z}$, where $\omega_z$
is the harmonic oscillator frequency along the tube. The peaks of the pair momentum
distribution are approximately given by $\pm |\tilde{k}^\uparrow_F-\tilde{k}^\downarrow_F|$,
and are the signature of the FFLO state.
}
\label{nk_pair}
\end{figure}

On the other hand, the \emph{pair} momentum distribution carries fingerprints of the
FFLO state. We define the pair two-body density matrix as
\begin{eqnarray}
\rho_\textrm{pair}(x,y)=  &&
\int \ud x_3 \cdots \ud x_N \Psi(x^\uparrow,x^\downarrow,x_3,\cdots,x_N) \times
\nonumber \\
&& \Psi(y^\uparrow,y^\downarrow,x_3,\cdots,x_N),
\label{pair_density_matrix}
\end{eqnarray}
where the pair $\{x^\uparrow,x^\downarrow\}$ is chosen by selecting $x^\uparrow$ and finding
$x^\downarrow$ as the position of the closest unlike-spin particle.
$x$ ($y$) is the center of mass of the $\{x^\uparrow,x^\downarrow\}$ ($\{y^\uparrow,y^\downarrow\}$) pair.
This is an unambiguous and generic way to define the pairs in a continuous system without introducing
explicitly a characteristic length. Once $\rho_\textrm{pair}(x,y)$ is computed,
the pair momentum distribution function $n_\textrm{pair}(k)$
is evaluated via Eq.~\ref{momentum_distribution}, with $\rho(x,y)$ replaced by $\rho_\textrm{pair}(x,y)$.
Also the normalization is based on the same formula as in Eq.~\ref{norm}, but with
$N$ replaced by $N_\textrm{pair}$, i.e. the number of pairs in the system.
The results for $n_\textrm{pair}(k)$ are presented in Fig.~\ref{nk_pair} for $g=20$, $N=200$,
and various polarizations. A clearcut signature of the FFLO state is the presence
of two peaks at $k=\pm | k^\uparrow_F-k^\downarrow_F|$, which signals a pairing
with non zero total momentum given by the difference between the Fermi surfaces of
up and down spin particles. Since the trap breaks the translational invariance,
the Fermi momenta are approximated by $\tilde{k}^\sigma_F = \frac{\pi}{2} \frac{N^\sigma}{L}$,
where $L$ is the effective spread of the cloud. The location of the peaks in Fig.~\ref{nk_pair}
follows closely the relation $\pm |\tilde{k}^\uparrow_F-\tilde{k}^\downarrow_F|$.

\begin{figure}[h] 
\centering
\includegraphics[angle=0, width=\columnwidth]{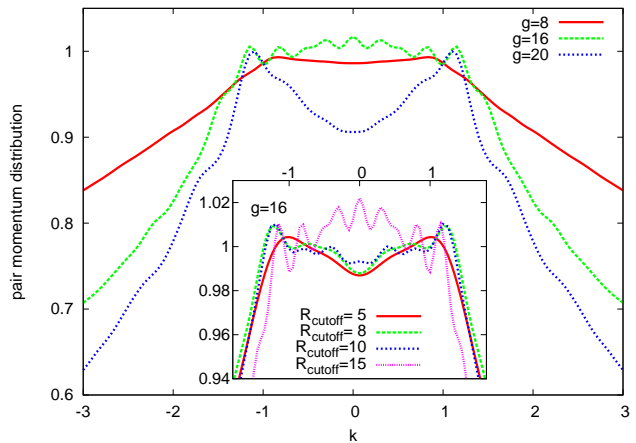}
\caption{(Color online) Pair momentum distribution
for $P=4\%$, $N=200$, at various couplings $g$.
The functions have been rescaled such that the height of their
peaks is around $1$ to make the comparison easier.
The inset displays the pair momentum distribution for $g=16$
computed from $\rho_\textrm{pair}(x,y)$
with $x,y \in [-R_\textrm{cutoff},R_\textrm{cutoff}]$.}
\label{pairnkvsg}
\end{figure}

Fig.~\ref{pairnkvsg} displays the $n_\textrm{pair}(k)$ dependence on $g$ at fixed
polarization $P=4\%$. At weak coupling ($g=8$) the peaks are short, and in the range
$[-\tilde{k}^\uparrow_F,\tilde{k}^\downarrow_F]$ the pair momentum distribution is almost flat.
On the other hand, at $g=20$ the two peaks at $\pm |\tilde{k}^\uparrow_F-\tilde{k}^\downarrow_F|$
are taller, and the atoms in the cloud have a quite strong FFLO character.
In the intermediate case analyzed here, namely $g=16$, the $n_\textrm{pair}(k)$ shows
a fluctuating behavior, with
peaks at $\pm |\tilde{k}^\uparrow_F-\tilde{k}^\downarrow_F|$
but also at $k=0$. By computing the Fourier transform (Eq.~\ref{momentum_distribution})
of $\rho_\textrm{pair}(x,y)$ constrained in the region with
$x,y \in [-R_\textrm{cutoff},R_\textrm{cutoff}]$
(and plotted in the inset of Fig.~\ref{pairnkvsg}), we proved that
the FFLO peaks come from the inner shell of the cloud, while
the central peaks come from the wings.
Indeed, around $g=16$ the system undergoes
the transition from fully polarized to fully paired (BCS) wings (see Fig.~\ref{polvsg}).
Both the FFLO pairing in the center and the BCS pairing in the outer shell
are weak. Going from the center to the edge, the crossover between the two
gives rise to this fluctuating pattern in the pair momentum distribution.

\subsection{Finite temperature}
\label{finite_temperature_results}

We performed finite temperature PIMC calculations for $g=8$, $16$, and $20$,
with $P=4\%$, and $N=200$.
From the DMC analysis and the TF results we know that for $g=8$
the phase separation is between an FFLO center and a fully polarized wing, while
for $g=16$ and $20$ the outer shell is fully paired. Here we study how the phase
separation evolves with the temperature.

\begin{figure}[h] 
\centering
\includegraphics[angle=0, width=\columnwidth]{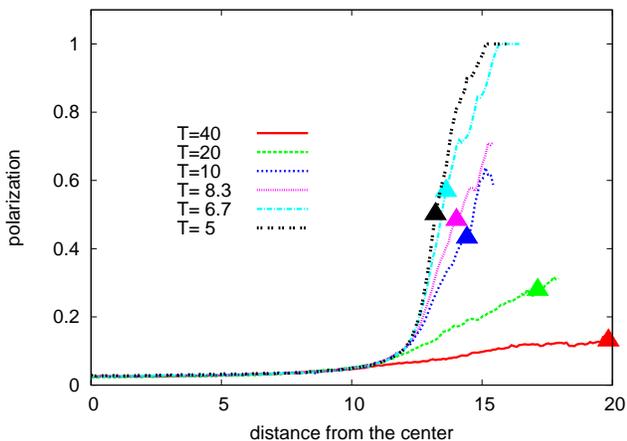}
\caption{(Color online) 
Local polarization profile for $N=200$, $P=4\%$, $g=8$, at various temperatures $T$
measured in units of $\hbar\omega_z$. For comparison, the central density
in the trap corresponds to an energy $\epsilon_F=\hbar^2/2m k_F^2\simeq 229$.
At this coupling the outer shell is fully polarized only at the two lowest temperatures
considered here, while it becomes partially polarized at a temperature in between $T=7$ and $T=8$.
The filled triangles indicate the polarization at the cloud radius 
which contains $99.9\%$ of atoms. 
The remaining $0.1\%$ ($0.2$ atoms) at the edge will be hard to detect in the experiment.} 
\label{pol_pimc_g8}
\end{figure}

In Fig.~\ref{pol_pimc_g8} we plot the local polarization profile for $g=8$.
At a temperature above $0.065 ~T^\uparrow_F$,
the fully polarized shell becomes
partially polarized with polarization progressively reduced
as the temperature is raised. At $T \simeq 0.40 ~T^\uparrow_F$ the polarization
is quite homogeneous throughout the cloud.

\begin{figure}[h] 
\centering
\includegraphics[angle=0, width=\columnwidth]{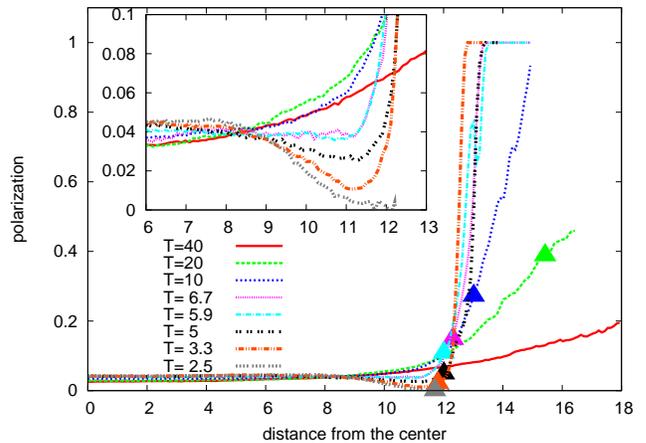}
\caption{(Color online) Local polarization profile for $N=200$, $P=4\%$, $g=16$,
at various temperatures $T$ measured in units of $\hbar\omega_z$.  For comparison, the central density
in the trap corresponds to an energy $\epsilon_F= \hbar^2/2m k_F^2 \simeq 259$. The inset magnifies
the region of the cloud around the polarization dip, where the pairing is the strongest.
For temperatures above $2.5$ the fully paired region starts breaking. When a pair is broken, the unpaired 
fermions feel the difference in the chemical potential
between the up and down species, resulting in the majority of up fermions 
at the edge of the trap, and a fully polarized outer shell.
For even higher temperatures, the cloud becomes partially polarized everywhere.
The filled triangles indicate the polarization at the cloud radius 
which contains $99.9\%$ of atoms.
}
\label{pol_pimc_g16}
\end{figure}

\begin{figure}[h] 
\centering
\includegraphics[angle=0, width=\columnwidth]{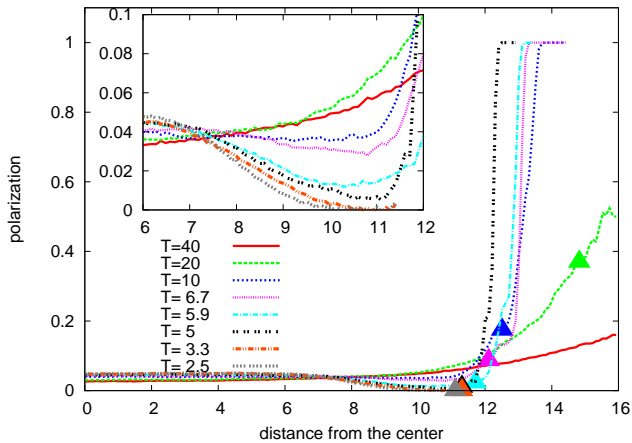}
\caption{(Color online) Local polarization profile for $N=200$, $P=4\%$, $g=20$,
at various temperatures $T$ measured in units of $\hbar\omega_z$.  For comparison, the central density
in the trap corresponds to an energy $\epsilon_F=\hbar^2/2m k_F^2\simeq 272$. The inset magnifies
the region of the cloud around the polarization dip, where the pairing is the strongest.
At this stronger coupling, the fully paired region is more stable, and starts breaking at a higher temperature 
(above $T=3.5$ is this case). The filled triangles indicate the polarization at the cloud radius 
which contains $99.9\%$ of atoms. 
}
\label{pol_pimc_g20}
\end{figure}

The finite temperature results at $g=16$ ($20$), reported in
Fig.~\ref{pol_pimc_g16} (\ref{pol_pimc_g20}), show a more
complicated pattern. Up to a temperature of $T \simeq 0.06
~T^\uparrow_F$ ($0.10 ~T^\uparrow_F$), the polarization profile
has a dip around $x=11.5$ ($10.5$), which evolves into a fully
paired region (with zero polarization) below $T \simeq 0.025
~T^\uparrow_F$ ($0.035 ~T^\uparrow_F$). When the dip is
present, the polarization in the outer part of the cloud
increases until a fully polarized state occurs at the very edge
of the cloud. This esternal region involves only $0.1\%$ of the
atoms in the cloud, and so it will be hard to detect in the experiment.
At higher temperatures, when the dip is gone, the
fully polarized edge becomes unstable. For $T \simeq 0.40
~T^\uparrow_F$ we have a situation analogous to the case with
$g=8$, with a partially polarized state homogeneously spread
over the cloud.

A feature common of all couplings is a crossover temperature $T_c$
above which the phase separation disappears and the whole cloud
becomes partially polarized. This temperature depends on the
polarization, since it scales roughly as
$\mu^\uparrow-\mu^\downarrow$, whereas its dependence on the
coupling is quite weak, at least until $g=20$. We found  $T_c
\simeq 0.065 ~T^\uparrow_F$ for $g=8$, and $T_c \simeq 0.10
~T^\uparrow_F$ for $g=20$ with $P = 4\%$. In case of fully
polarized wings, one can define another critical temperature
based on the stability of the pairs at the edge of the cloud.
The fermions dissociate into a partially polarized fluid if the
temperature is raised above a threshold proportional to $g^2$,
as it depends on the binding energy of each pair.

As in the DMC calculations, the PIMC simulations suffer from
the ergodicity problem mentioned before. This is an issue of
the strictly one dimensional Monte Carlo sampling, where the
exchange between particles or pairs is suppressed at strong
coupling. For this particular system, we found that this issue
is less severe in the DMC simulations than in the PIMC ones. 
Improved moves or working in the grand
canonical ensemble could alleviate or solve the problem. Here,
however, we could afford simulations with $N=200$ at quite
strong coupling, namely up to $g=50$ in the DMC, and $g=20$ in
the PIMC framework. Therefore our analysis has not been
limited, since we were able to study the intermediate-strong
coupling regime, where the fully paired region shows up at
not-so-small polarizations.

\section{Conclusions}
\label{conclusions}

In this paper we have studied the properties of a system
of 1D trapped fermions interacting via an
attractive contact potential, by performing unbiased DMC and PIMC simulations.
We showed that the trap induces the system to phase separate, in accordance
to previous calculations done with the TF approximation.\cite{orso,liu_hu_drumm_prl}
We compared the TF solution with our
exact results, by studying their dependence on the coupling and system size.
We found that the DMC phase boundaries are in a good
agreement with the TF approximation for large number of particles
($N \gtrsim 100$) even at strong coupling. This is reasonable
since the TF approach is applicable when $N \gg 1$, and
the size of the cloud is large compared to the axial oscillator length.
However, the local density approximation requires also
a slow variation of the mean interparticle spacing,
namely $\partial_x n(x) / n(x)^2 \ll 1$, condition which is not fulfilled
by the system particularly at the edge of the cloud,
where the TF solution slightly overestimates
the size of the outer fully paired region.
For smaller systems we found that the agreement deteriorates. Indeed,
at small $N$ a partially polarized state extends until the
edge of the cloud even at intermediate coupling and small spin imbalance,
in contrast with the TF approximation, which predicts a phase separation
with fully paired wings.

As pointed out in a number of previous works\cite{batrouni,tezuka,feiguin,rizzi}
the pair momentum distribution reveals
the signature of the FFLO inner shell with two symmetric peaks
centered at $\pm | k^\uparrow_F - k^\downarrow_F |$.
It could be measured in the experiment by quickly removing the confining lattice and
sweeping the magnetic field to the BEC side of the Feshbach resonance.
This would bind the pairs into molecules,
whose momentum distribution could be measured
by means of time-of-flight imaging techniques.\cite{greiner}
The tails of the one-body momentum
distribution strongly depend on the interaction strength and
decay as $1/k^4$, due to the non analyticity
of the exact wave function as a consequence of the delta function interaction.
This could be another interesting quantity to measure in the experiment,
although it does not carry information on the actual pairing in the system, but only
on the effective strength.

The finite temperature PIMC calculations done with $N=200$, $P=4\%$, and three different
couplings ($g=8$, $16$, and $20$) show a crossover from a phase separated state to a partially
polarized state at $T \simeq 0.065 ~T^\uparrow_F$ for $g=8$,
and $T \simeq 0.10 ~T^\uparrow_F$ for $g=20$.
For the strongest coupling analyzed with this method ($g=20$),
the fully paired region disappears already at
$T \simeq 0.035 ~T^\uparrow_F $, leaving a dip in the polarization density profile which
becomes shallower and shallower as the temperature increases.
This quite low temperature (the experiments will access temperatures down to
$T \simeq 0.05 ~T^\uparrow_F$) puts some limitation on the visibility
of the fully paired phase at weak-intermediate coupling. One has to
go to stronger coupling to make the pairs in the
outer shell stable up to higher temperatures, being the binding energy
proportional to $g^2$.

We discussed the ergodicity issues in the Monte Carlo sampling at strong coupling,
and we identified the problem with the difficulty of unpaired fermions to pass through
strongly bound pairs. The same slowing down in the equilibration might occur
also in the experiment for ultra thin traps ($N \hbar \omega_z \ll \hbar \omega_\perp$
and $k_B T \ll \hbar \omega_\perp$).
On the other hand, at weak coupling there are few particles involved in the fully paired
region, which is very narrow, and the signal coming from this phase should be very difficult
to detect in the experiment.
We suggest that a good spot to detect the transition from
the fully paired to the fully polarized state is
for \emph{effective} couplings  $g/\sqrt{N}$  in the range $3-5$,
where the system features a fully paired region up to
$10\%$ of spin imbalance, the size of the two phase separated states is
quite large, and the equilibration is not hard to reach.

The information coming from this paper could be combined with the recent work
by Parish \emph{et al.},\cite{huse} who studied the quasi one dimensional effects coming
from the array of tubes in the transverse direction by means of a tight-binding model,
in order to find out the parameters of the Hamiltonian which maximize the visibility
of the FFLO state. In principle, one can extend
the present work by explicitly including an array of tubes in the QMC simulations
to study quasi one dimensional effects present in the experimental setup.

\acknowledgments
We thank R. Hulet, D. Huse, A. Minguzzi, and D. Trinkle for fruitful discussions.
We acknowledge support in the form of the NSF grant DMR-0404853
and from ARO Award W911NF-07-1-0464 with funds from the DARPA
OLE Program  and computer resources from NCSA .

\bibliographystyle{h-physrev}
\bibliography{trapped_1d}

\end{document}